# Tight binding description of the STM image of molecular chains

Yoel Calev,[a] Hezy Cohen,[b] Gianaurelio Cuniberti,[c] Abraham Nitzan,[a] and Danny Porath[b]

[a] School of Chemistry, Tel Aviv University, Tel Aviv, 69978, Israel

[b] Institute of Chemistry, The Hebrew University, Jerusalem, 91904, Israel

[c] Institute for Theoretical Physics, University of Regensburg, D-93040 Regensburg, Germany

## Abstract

A tight binding model for scanning tunneling microscopy images of a molecule adsorbed on a metal surface is described. The model is similar in spirit to that used to analyze conduction along molecular wires connecting two metal leads and makes it possible to relate these two measurements and the information that may be gleaned from the corresponding results. In particular, the dependence of molecular conduction properties along and across a molecular chain on the chain length, intersite electronic coupling strength and on thermal and disorder effects are discussed and contrasted. It is noted that structural or chemical defects that may affect drastically the conduction along a molecular chain have a relatively modest influence on conduction across the molecular wire in the transversal direction.

# 1. Introduction

Electronic conduction through individual molecules connecting two metal leads was suggested a long time ago as a mechanism for molecular rectification,[1] but was realized only years later with the invention of the scanning tunneling microscope (STM).[2] More recently, molecular conduction has been studied using other setups such as break junctions, fixed lithographically prepared closely separated electrodes[3] and electro-migration controlled leads.[4] In both experimental and theoretical studies the junction conduction behavior as expressed by its current ($I$) – voltage ($\Phi$) characteristic is of course a central issue. However the dependence of this behavior on molecular and environmental properties is also of fundamental and practical importance. Indeed, the dependence of the conductance of a molecular bridge on its length, on the molecule-lead binding, the intersite coupling within the bridge, the molecular periodic vs. disordered structure, effects of symmetry and of chemical substitution, as well as thermal and dephasing effects including transitions from coherent tunneling to incoherent hopping transport, was studied in the past decade using generic tight binding molecular models.

In principle, molecular conduction can be studied in different configurations of relative leads-molecule positions and orientations. In what follows we discuss in particular two such configurations: (*a*) conduction along a molecule connecting two metal leads, and (*b*) a scanning tunneling microscope experiment in which a molecular chain lies flat on a surface of a conducting substrate (see Fig 1). In the absence of thermal effects the molecular conduction at energy $E$ can be described in both cases by the Landauer formula[5,6]

$$g(E) = \frac{e^2}{\pi\hbar}\mathcal{T}(E) \quad (1)$$

where the transmission coefficient is given by

$$\mathcal{T}(E) = 4Tr\left\{\Gamma^{(L)} G_M \Gamma^{(R)} G_M^\dagger\right\} \quad (2)$$

Here $G_M$ is the molecular Green's function and $\Gamma^{(K)} = (i/2)\left(\Sigma^{(K)\dagger} - \Sigma^{(K)}\right)$, $K = R, L$ where $\Sigma^{(L)}$ and $\Sigma^{(R)}$ are the self energy matrices associated with the couplings of the molecule to the "left" and "right" leads, respectively. $G_M(E)$ is given in terms of the molecular Hamiltonian and of these self energies by

$$G_M(E) = \left[E - H_M - \left(\Sigma^{(L)} + \Sigma^{(R)}\right)\right]^{-1} \equiv \left[E - H'_M + i\left(\Gamma^{(L)} + \Gamma^{(R)}\right)\right]^{-1} \qquad (3)$$

Here $H'_M = H_M + \text{Re}\left(\Sigma^{(L)} + \Sigma^{(R)}\right)$ is the molecular Hamiltonian renormalized by the molecule-leads coupling.[7] The zero bias conduction is given by $g(E_F)$ where $E_F$ is the leads' Fermi energy, while the current at finite bias voltage $\Phi$ can be calculated from

$$I(\Phi) = \frac{e}{\pi\hbar} \int_{-\infty}^{\infty} dE\, \mathcal{T}(E,\Phi)\left(f\left(E + \frac{e\Phi}{2} - E_{f_L}\right) - f\left(E - \frac{e\Phi}{2} - E_{f_R}\right)\right) \qquad (4)$$

where $f(E)$ is the Fermi function

$$f(E) = \left(1 + e^{-E/k_B T}\right)^{-1} \qquad (5)$$

and $k_B$ and $T$ are the Boltzmann constant and the system temperature respectively. Note that in Eq. (4) the transmission coefficient $\mathcal{T}$ depends in principle on the bias voltage $\Phi$ and on the way it falls along the molecule.[6]

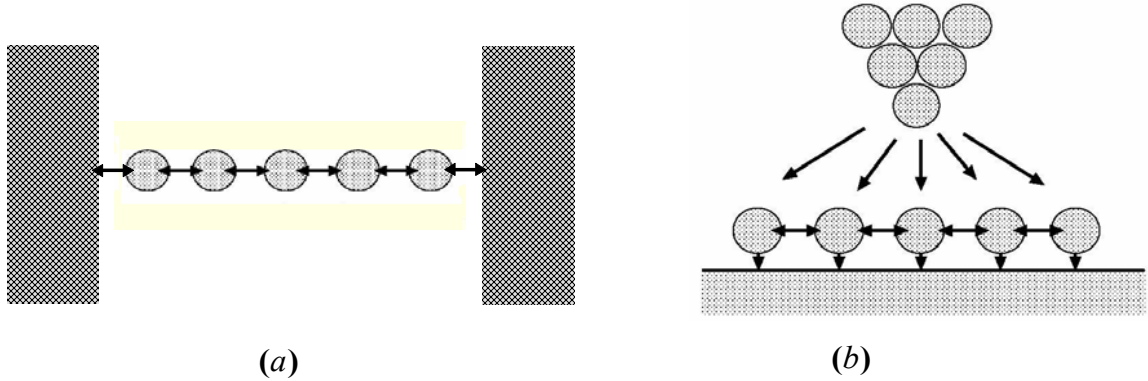

(a)    (b)

Fig. 1. Two configurations of a molecular conduction junction: (a) A molecular wire connecting between two metal leads. (b) Conduction across a molecule using an STM tip above a flatly adsorbed molecule.

Fig. 1 depicts simple models of a molecular chain connecting two metal leads in the two configurations mentioned above. Fig. 1a shows a molecular chain connected longitudinally between two metal leads. Fig. 1b shows an STM configuration in which the same molecule lies flat on the substrate surface and is being scanned with an STM tip. There are a few basic differences between the two configurations. In the first, charge carriers are injected from one lead into the molecule at one of its ends, travel all along the molecule to its other end and pass into the other lead. The molecule-metal bonding in this type of experiment is usually strong, implying relatively small potential barriers for the molecule-metal electron transport. In the STM configuration, a metal lead (the STM tip) approaches from above to any point along the molecule, charge carriers tunnel towards that point and find their way through the molecule to the substrate. Here the molecule-metal contact is usually non chemical, implying a large barrier for electron injection (relative to the case of chemical bonding) into and out of the molecule, in particular on the tip side.

Another important difference between the two configurations is their expected sensitivity to impurities and defects along the molecular chain. In configuration *a* the conduction depends strongly on the electronic coupling between the consecutive sites. Therefore even a single defect in the chain may dramatically affect the observed conduction. In configuration *b* the STM tip can approach any site along the molecule and conduction takes place essentially through this site. Therefore the sensitivity to impurities and defects is relatively small. In both configurations the existence of a surface under the molecule may affect the molecular electronic structure, the intersite electronic coupling and the defect distribution.

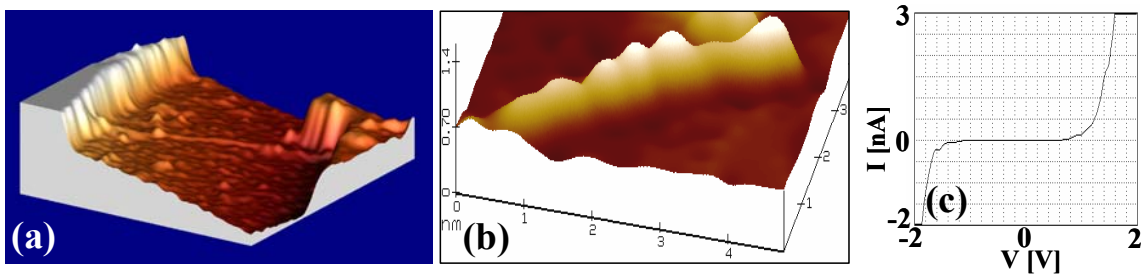

Fig. 2 – (a) An AFM image of a single DNA molecule, connected to two gold electrodes, lying flat on a $SiO_2$ surface. (b) A single DNA molecule lying on a metal surface, as imaged using STM ($I$=0.2 nA, $V_b$=2.5 V) and (c) an example of a current-voltage curve measured on a DNA molecule using STM at room temperature (we note that details of consecutive curves were not reproducible at this temperature).

Fig. 2 shows actual realizations of these situations. Fig 2a shows an atomic force microscope (AFM) image of a single DNA molecule connected to two metal electrodes, lying flat on a $SiO_2$ surface. Fig 2b shows a short DNA molecule lying flat on a metal surface, as imaged using STM. The high "peaks" in the molecule are probably the DNA base-pairs. Note that when the molecule is lying flat on the surface it is likely that its helical structure will be bent and deformed, possibly imposing bends and defects and consequently high potential barriers for electrical transport. Fig. 2c depicts an example of room temperature current-voltage curve for a DNA molecule adsorbed on a gold surface and scanned with an STM Pt-Ir tip.[8]

In the case where coherent tunneling is the dominant mechanism for conduction in configurations *a* and *b*, Eqs. (1) and (4) apply to both. Assuming that the molecular electronic structure is essentially the same in the two configurations (an approximation that may hold true because both observation are made for a molecule lying on a supporting surface) the main difference between the two cases arises from the self energy matrix $\Sigma$. Adopting a simple nearest neighbor tight binding model for the molecular Hamiltonian so that in a local representation with one orbital per site

$$H_M = \begin{pmatrix} E_M & V_M & 0 & 0 \\ V_M & E_M & \ddots & 0 \\ 0 & \ddots & \ddots & V_M \\ 0 & 0 & V_M & E_M \end{pmatrix} \tag{6}$$

the self energy matrices associated with the leads in case *a* (disregarding the coupling to the leads of all but the nearest molecular sites) are of the form

$$\Sigma_a^{(L)} = \begin{pmatrix} \sigma_L & 0 & \cdots & 0 \\ 0 & 0 & & \\ \vdots & & \ddots & \\ 0 & & & 0 \end{pmatrix} ; \quad \Sigma_a^{(R)} = \begin{pmatrix} 0 & 0 & \cdots & 0 \\ 0 & 0 & & \\ \vdots & & \ddots & \\ 0 & & & \sigma_R \end{pmatrix} \tag{7}$$

while in case *b* the self energy due to the substrate is

$$\Sigma_b^{substrate} = \begin{pmatrix} \sigma_1 & \sigma_2 & \sigma_3 & \ddots & \\ \sigma_2 & \sigma_1 & \sigma_2 & \ddots & \ddots \\ \sigma_3 & \sigma_2 & \ddots & \ddots & \sigma_3 \\ \ddots & \ddots & \ddots & \sigma_1 & \sigma_2 \\ & \ddots & \sigma_3 & \sigma_2 & \sigma_1 \end{pmatrix} \tag{8}$$

In fact, below we approximate this by a diagonal form, disregarding $\sigma_j$ for j>1. The rational for this approximation is that non-diagonal terms in $\Sigma_b^{substrate}$ are expected to be small if the characteristic distance between the effective molecular sites is considerably larger than the electronic screening length that characterizes the substrate. The corresponding matrix for the tip depends on the tip's position and is discussed below.

An important observation is that conduction associated with both configurations *a* and *b* arises from the same molecular Hamiltonian $H_M$, i.e. depend on the same essential parameters: the barrier height $(E_M - E_F)$ and the intersite coupling $V_M$. This remains true when the conduction is dominated by thermal activation and hopping along the bridge, where transport depends also on the temperature and thermal relaxation rates that together with the Hamiltonian parameters determine the activation probability and the hopping rates. Measuring and computing the molecular current-voltage characteristics in both configurations *a* and *b* can therefore provide, in principle, an important consistency check on any theoretical interpretation of the observed behavior. However, to carry out such a program we would need a good characterization of the molecule-lead coupling in both configurations.

In the present paper we undertake the considerably simpler goal of comparing, within the same tight binding model of the molecular Hamiltonian, the conduction properties of a molecular chain in configurations *a* and *b*. In particular we focus on the STM configuration *b* since many studies of this model for conduction in configuration *a* were already carried out. (see, eg. Refs. 4, 9) We focus on the same generic issues that were subjects of these studies: the dependence of observed signals on molecular parameters (chain-length and intersite coupling) and the effects of structural disorder and thermal relaxation.

## 2. The Model

We focus on the STM configuration *b* but in order to keep our notation uniform we will continue to use the labels *L* and *R* for the leads. For specificity we will use the label *L* for the tip and the label *R* for the substrate. The junction Hamiltonian can be written as a sum of the Hamiltonians of the free molecule, $H_M$, the tip, $H_L$, the substrate, $H_R$, and their mutual couplings:

$$H = H_M + H_L + H_R + V_{LM} + V_{RM} \tag{9}$$

For simplicity we disregard the direct electronic coupling between tip and substrate. The effective Hamiltonian that describes dynamics in the molecular subspace is

$$H_M^{eff} = H_M + \Sigma^{(L)} + \Sigma^{(R)} = H_M' - i\left(\Gamma^{(L)} + \Gamma^{(R)}\right) \qquad (10)$$

where $\Sigma^{(K)}$ ($K=L, R$) are the self energies discussed above, and $\Gamma^{(K)}$ are minus their imaginary parts. The real parts of these self energies constitute an energy renormalization of the molecular Hamiltonian, cf Eq. (3). This renormalized Hamiltonian $H_M'$ is represented in our model by an $N$-site nearest neighbor tight-binding form

$$H_M' = \sum_{n=1}^{N} E_n |n\rangle\langle n| + \sum_{n=1}^{N-1} \left(V_M |n\rangle\langle n+1| + V_M^* |n+1\rangle\langle n|\right) \qquad (11)$$

The local basis $\{|n\rangle\}$ used here is assumed to be orthogonal. An "ordered" molecular chain will be characterized by the same energy, $E_n = E_M$, for all sites, while site disorder will be represented by sampling the site energies $E_n$ from some random distribution.[10] Fig. 3 shows the parameters needed to calculate the STM current signal in this model.

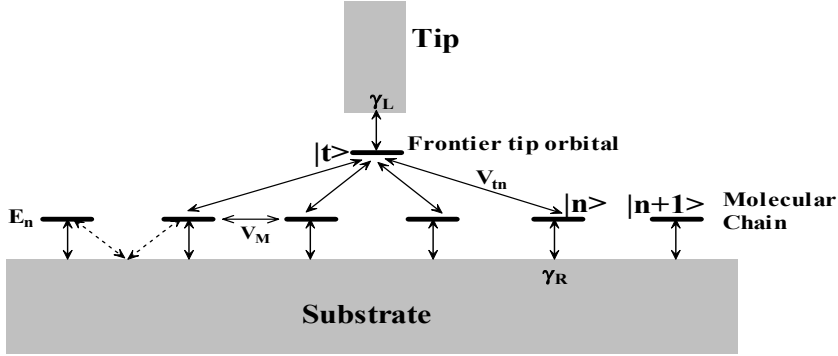

fig. 3 A schematic presentation of the electronic states and coupling parameters that characterize the model used in this work. Shaded areas correspond to the continuum of quasi-free electronic states on the tip and the substrate. $|n\rangle$ ($n=1,\ldots,N$) are states localized on the molecular segments that, together with their energies $E_M$ and nearest-neighbor interstate coupling $V_M$, define the molecular bridge. $|t\rangle$ is the tip "frontier" orbital whose coupling to the rest of the tip is characterized by the damping parameter $\gamma_L$. The local molecular states $|n\rangle$ are coupled to the tip via their coupling $V_{tn}$ to this orbital, and their coupling to the substrate is expressed by the damping parameter $\gamma_R$.

Consider now the damping matrices, $\Gamma^{(L)}$ associated with the molecule-tip interaction, and $\Gamma^{(R)}$ that results from the molecule-substrate coupling. For an "ordered" chain with identical repeat units the matrix $\Gamma^{(R)}$ is expected to be of the form (8), where non diagonal terms result from the interactions between different molecular units with the same lead modes. If the spatial distance between these units is large relative to the typical screening length these non diagonal terms will be small, and are disregarded in what follows. Effectively this amounts to assuming that each molecular site in Fig. 3 is coupled to its own substrate. Under this assumption we get

$$\Gamma^{(R)} = \gamma_R \mathbf{I} \qquad (12)$$

where $\mathbf{I}$ is a unit matrix of order $N$ – the number of molecular sites.

A reasonable model for $\Gamma^{(L)}$ may be obtained by assuming that the molecule-tip coupling is mediated by a single atomic "frontier" orbital $|t\rangle$ at the tip edge. This is a local atomic orbital that is coupled strongly to the rest of the tip: an excess electron placed in this orbital will delocalize on the tip on a timescale of order $\gamma_L^{-1}$, where $\gamma_L$ is of the order of the tip conduction bandwidth. In practical situations $\gamma_L$ is much larger than the energy mismatch between the injection energy $E$ (of order of the Fermi energy) $E_F$ and the (zero order) energy $E_t$ of tip frontier orbital $|t\rangle$. Therefore the decay matrix $\Gamma^{(L)}$ from the molecular bridge to the tip will be of the order

$$\Gamma^{(L)}_{n,n'} \approx \frac{V_{n,t} V_{t,n'} \gamma_L}{(E-E_t)^2 + (\gamma_L)^2} \approx \frac{V_{n,t} V_{t,n'}}{\gamma_L} \qquad (13)$$

A rough estimate of the coupling $V_{t,n}$ between the tip frontier orbital and molecular site orbitals may be obtained by taking both as the lowest states of square potential wells of spatial widths $l_M$ separated by a barrier of height given (relative to the Fermi energy) by the metal work-function $W_F$ and width equal to the spatial separation $d_{tn}$ between the tip site $t$ and the molecular site $n$. $V_{t,n}$ is identified as half the ground state energy splitting in this double-well structure, given approximately by[11]

$$V_{t,n} = \frac{\hbar^3 \pi^2 \sqrt{2mW_f}}{m^2 l_M^3 \left(\frac{\hbar^2 \pi^2}{2m l_M^2} + W_f\right)} e^{-\frac{d_{tn}}{\hbar}\sqrt{2mW_f}} \qquad (14)$$

Next we discuss our choice of model parameters, keeping in mind that our primary goal is the comparison between conduction along a molecular bridge and across it, not obtaining absolute numbers. Placing the Fermi energies of the metal leads at zero, we take $E_M = 0.5$ eV and $V_M = 0.1$ eV as the parameters associated with the molecular electronic structure. These are of the order used to model conduction along DNA molecules. We choose the width $\gamma_R$ of Eq. (12) as $\gamma_R$ = 0.012 eV (~100cm$^{-1}$), about a tenth of level widths associated with chemical adsorption. $\gamma_L$ in Eq. (13) and $W_f$ in Eq. (14) (orders of metal bandwidth and workfunction) are taken 5eV. The length parameter $l_M$ in Eq. (14) (order of an orbital spatial size) is taken as 0.2nm and the molecular-site - tip distance $d_{t,n}$ is calculated by assuming that the tip-molecule distance is 0.6nm and that the distance between nearest-neighbor molecular sites is 0.34nm (the distance between neighboring base pairs in B-DNA).

Finally, to calculate conduction along the molecular chain (configuration *a*) we use $\Gamma^{(K)} = (i/2)\left(\Sigma^{(K)\dagger} - \Sigma^{(K)}\right)$, $K = R, L$ from Eq. (7), with the choice $\gamma_K = -\text{Im}(\sigma_K) = 0.2 eV$, a typical level broadening associated with chemisorption. Furthermore, we examine two models for the potential bias distribution along the molecule: For a total bias $\Phi$ one model is a linear drop, $E_n = E_M + \Phi/2 - n\Phi/(N+1)$ ; $n = 1, 2, ..., N$. The other assumes that the potential drops linearly between one lead and sites *n*=2 on one side, the other lead and the site *n*=*N*-1 on the other, and is constant on the rest of the molecule between sites 2 and *N*-1.

Eqs. (9)-(14) together with the choice of parameters outlined above fully define our model. In simple limits of this model the zero bias ($\Phi \to 0$) conductance $g(E_F)$ can be written in convenient analytical forms. Closed forms for the end-to-end linear chain (Fig. 1.a) conductance exist in the literature[12]. For the STM configuration it is also possible to obtain analytical results if we assume an STM tip in coupled only to one atom, *k*, of a *N*-site molecular chain (that is $\left(\Sigma_L\right)_{nn'} = -i\Gamma^{(L)}_{kk}\delta_{nk}\delta_{n'k}$).[13] By denoting as $G^{\text{surf}} = \left(E - H_M - \Sigma_R\right)^{-1}$ the chain Green function dressed by the surface self-energy, we obtain

$$g_k = \frac{e^2}{\pi\hbar} \frac{4\Gamma^{(L)}_{kk}\Gamma^{(R)}_{kk}}{\left|G^{\text{surf}-1}_{kk} + i\Gamma^{(L)}_{kk}\right|^2} \sum_{n=1}^{N} \left|\frac{G^{\text{surf}}_{nk}}{G^{\text{surf}}_{kk}}\right|^2 \qquad (15)$$

note that in this case the *k* dependence represents here only a finite size effect (it is washed out in the large *N* limit). Eq. (15) can be made explicit in the case where only the main diagonal of the matrix (8) is retained, with all elements equal, $\Gamma^{(R)}_{kk} = \gamma_R$.[14] In this case $G^{\text{surf}}$ may be written explicitly in terms of Chebyshev polynomials $U_n(x)$ of the second kind[15], with a complex scaled energy argument $x = (E_F - E_M + i\gamma_R)/(2V_M)$. In particular, for the central position $k = 0$ in a chain with an odd number of segments, $N=2\nu+1$, we can write

$$g = \frac{e^2}{\pi\hbar} \frac{4\Gamma_{kk}^{(L)}\gamma_R}{\left|V_M \frac{U_{2\nu+1}(x)}{U_\nu^2(x)} + i\Gamma_{kk}^{(L)}\right|^2} \left(1 + 2\sum_{n=1}^{\nu}\left|\frac{U_{\nu-n}(x)}{U_\nu(x)}\right|^2\right) \qquad (16)$$

Of particular interest is the dependence on the intramolecular coupling $V_M$. It is easy to see that $g$ approaches the Breit-Wigner result for the conductance through a single site at small $V_M$'s

$$g \xrightarrow{V_M \to 0} \frac{e^2}{\pi\hbar} \frac{4\Gamma_{kk}^{(L)}\gamma_R}{(E_F - E_M)^2 + (\Gamma_{kk}^{(L)} + \gamma_R)^2} \qquad (17)$$

(with the maximum obtained on resonance). In the opposite limit of large $V_M$ the denominator of Eq. (16) becomes relevant. Using

$$\frac{U_{2\nu+1}(x)}{U_\nu^2(x)} = \begin{cases} \sim x & (\nu \text{ even}) \\ \sim x^{-1} & (\nu \text{ odd}) \end{cases} \qquad (18)$$

we see that in this limit $g$ oscillates between 0 and a finite value for chains with odd and even $\nu$, respectively.[16]

Below we use the model described above to analyze basic properties of conduction along and across molecular chains.

## 3. The effect of coupling to the thermal environment

Eqs (1) - (4) result from a model of coherent elastic transport that has to be modified when dephasing and thermal relaxation effects become important. Coupling to the thermal environment may result in destruction of coherence, lead to inelastic contributions to the tunneling flux and can open a new, activated channel of conductance. All these processes may be described by the transmission function $\mathcal{T}(E_{in}, E_{out})$, which depends on both the incident ($E_{in}$) and the outgoing ($E_{out}$) energies that may now be different from each other. The total transmission probability at energy $E$

$$\mathcal{T}(E) = \int dE_{out} \mathcal{T}(E, E_{out}) \qquad (19)$$

is the analog of the corresponding quantity that appear in Eqs. (1) and (4), however these equations are not in general valid in the presence of thermal interactions. A proper description of transport in this case is provided in terms of the reduced molecular density matrix obtained by tracing out the environmental degrees of freedom from the equations of motion. For weak coupling between the system and the thermal environment this leads to a Redfield-type equation[17,18] for the molecular density matrix. In the present application we use a variant of the steady state scattering procedure of Segal and Nitzan[19,20] in order to calculate $\mathcal{T}(E)$. As in that work the total Hamiltonian (9) is supplemented by the thermal bath Hamiltonian $H_B$ and the molecule-bath interaction $H_{MB}$

$$H_{MB} = \sum_{n=1}^{N} F_n |n\rangle\langle n| \qquad (20)$$

where $F_n$ are operators of the thermal bath taken to satisfy $\langle F_n \rangle = 0$ and

$$\int_{-\infty}^{\infty} dt e^{i\omega t} \langle F_n(t) F_{n'}(0) \rangle = \begin{cases} C_T \delta_{n,n'} & ; \omega \geq 0 \\ e^{-\hbar|\omega|/k_B T} C_T \delta_{n,n'} & ; \omega < 0 \end{cases} \qquad (21)$$

Eq. (21) is a simple model constructed in accordance with the detailed balance relation

$$\int_{-\infty}^{\infty} dt e^{i\omega t} <F_n(t) F_n(0)> = e^{\hbar\omega/k_B T} \int_{-\infty}^{\infty} dt e^{i\omega t} <F_n(0) F_n(t)> \qquad (22)$$

Here $T$ is the temperature, $k_B$ - the Boltzmann constant and the parameters $C_T$ and $\tau_c$ characterize the molecule-bath coupling and bath correlation time, respectively. The variation from the procedure of Ref. [19,20] is[21] to divide the effective molecular Hamiltonian (10) to its hermitian and anti-hermitian parts, taking the latter as part of the interaction, together with $H_{SB}$, so that diagonalization in the basis of the 'unperturbed' problem does not require the use of different left and right eigenvectors. The final result of this calculation is the transmission function $\mathcal{T}(E_{in}, E_{out})$ and the overall transmission at energy $E$, Eq. (19). It should be emphasized that conduction through a molecule connecting two metal leads cannot be simply described by these transmission functions alone, because the Fermi distributions of electronic populations in the bridge affect the transmission in a non-trivial way (see Ref. 6, Chapter 8).

A proper description can be obtained using the non-equilibrium Green's function formalism. Here, for simplicity, we disregard this complication, and limit ourselves to analyzing thermal effects on the transmission function (19) itself.

## 4. Results and discussion

Fig. 4 shows the current-voltage characteristics, calculated from Eq. (4) at room temperature, for configurations *a* (along the molecule) and *b* (across the molecule) for a molecule with $N=7$. In configuration *b* the tip is placed above the central molecular site. Our choice of reasonable molecular parameters gives current-voltage dependence that fall within the range of observed behaviors in both types of measurements.

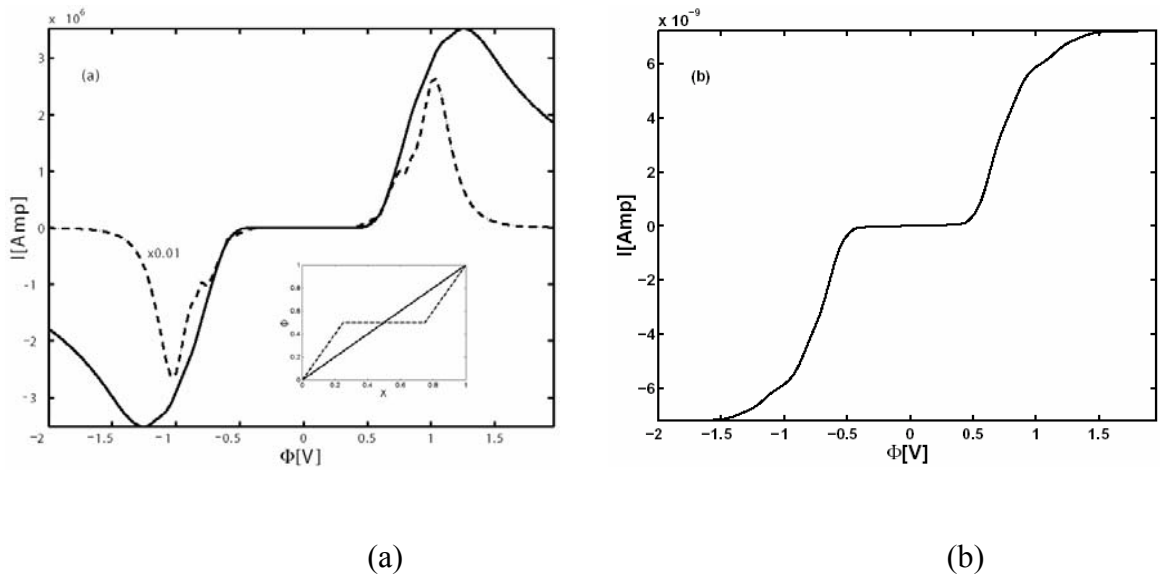

(a)  (b)

Fig. 4

Current vs. voltage in (a) Configuration *a* and (b) Configuration *b*, for an N=7 model. The parameters for this calculation are given in Section 2. In (a) the dashed line is the result obtained for the linear potential drop model and it is scaled by a factor 100 to fit into the current window shown. The full line results from a model in which the potential is assumed to drop linearly between the leads and the sites n=2 (6) and to be constant on the interior of the molecular chain between sites 2 to 6 (see inset). In (b) the tip is above the center molecular site.

Fig. 5 shows the dependence of the zero bias conduction $g(E_f)$ on the intersite molecular coupling $V_M$. Figs. 5*a* and 5*b* show the conduction in configurations *a* and *b* respectively for the 7-site molecule. Fig 5c show similar results for configuration *b* for the *N*=7 and *N*=100. Several features in these results are noteworthy. First, conduction along the molecule (configuration *a*) vanishes when $V_M$=0 as it obviously should. As $V_M$ increases the molecular levels change, and for $V_M$>0.25eV (for $E_M$=0.5eV) levels of the molecular "conduction band" (of width ~$4V_M$) becomes resonant with the Fermi energy whereupon conduction increases. Second, the existence of a "gap" about $V_M$=0, for the same reason, is seen also in configuration *b*, although the conduction in this case does not vanish when $V_M$ = 0. The dependence on molecular length in this case is small (see also Fig. 6). Third, the structure seen in *g* beyond this $V_M$=0.25eV reflects the discrete nature of the molecular states. Finally, it is interesting to note that while in configuration *a* the conduction dependence on $V_M$ is symmetric under sign inversion, in configuration *b* the $V_M$ dependence shows asymmetry about $V_M$=0. This phenomenon results from the fact the conduction in configuration *b* is affected by interference of contributions from several pathways, and therefore contain contributions that depend on the sign (in fact on the phase) of $V_M$.

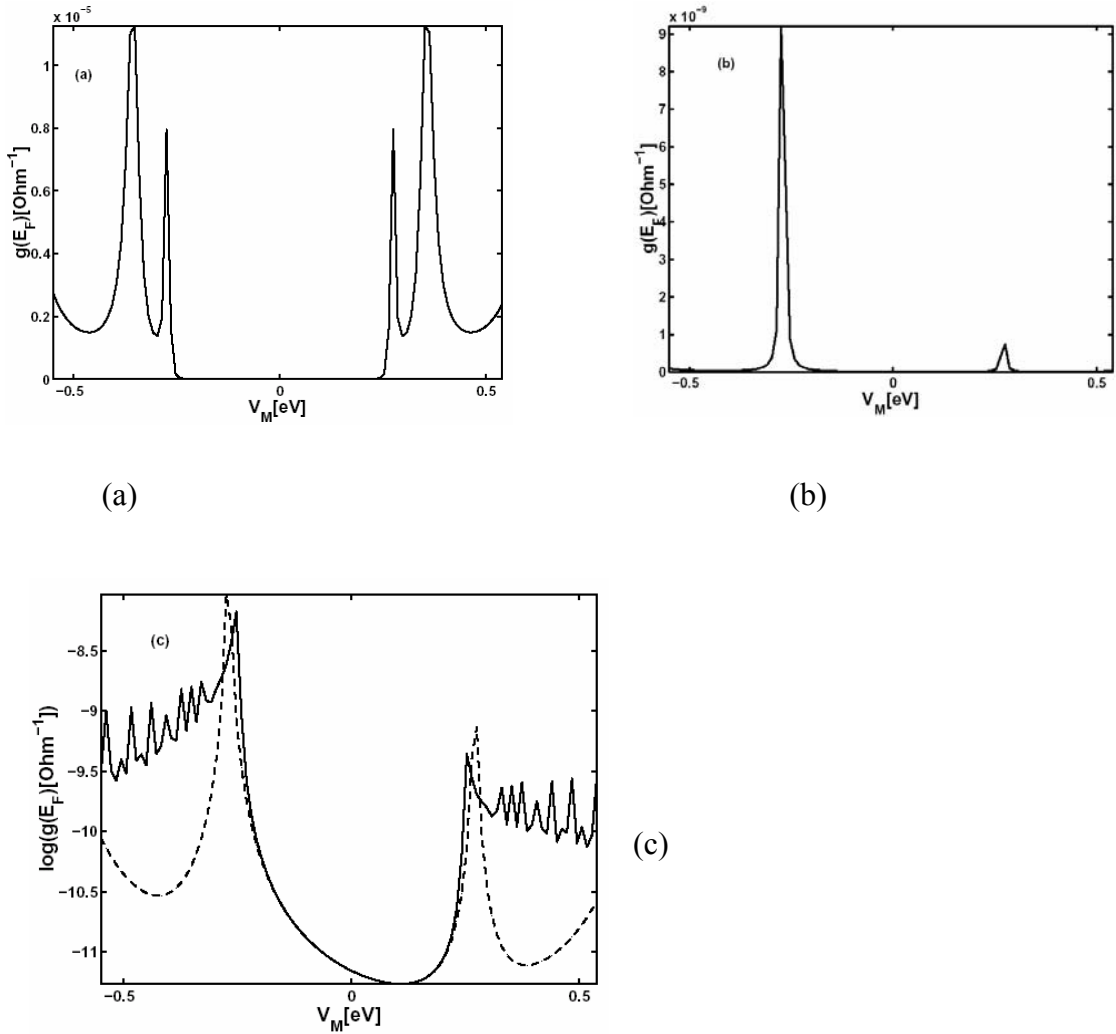

(a)

(b)

(c)

FIG. 5

Zero bias conduction as a function of intersite coupling $V_M$. (a) Conduction along the molecular axis (configuration *a*) for an *N*=7 model. (b) Conduction across the molecule (N=7) when a tip is above the central molecular site. (c) Same as (b), on a logarithmic scale, for models with *N*=7 (dashed line) and *N*=100 (full line).

The dependence of conduction along a molecular wire on the length of the wire length is a central attribute of a single molecule junction, similar in the scope of its implications to the dependence of bridge assisted molecular electron transfer rates on the bridge length. Two modes of behavior, tunneling and hopping, were found to yield vastly different dependence on the wire length.[22,23] Conductance dominated by coherent tunneling depends exponentially on the wire length (expressed in terms of the number of sites *N* along the bridge) according to

$$g = g_0 e^{-\beta N} \tag{23}$$

where the exponential damping parameter *β* is typically found in the range 0.5…1.5. When conduction is dominated by thermal activation onto the bridge and hopping along the bridge the length dependence is qualitatively different, having the form

$$g = (\alpha_1 + \alpha_2 N)^{-1} \tag{24}$$

These modes of behavior, as well as the crossover between them were discussed theoretically[22] and observed experimentally[24] for conduction along the molecular chain. For conduction across the chain, configuration *b*, the dependence on chain length does not have this intrinsic and practical importance; still, it is of interest to examine it while comparing the two junction configurations. Fig 6 shows the zero bias conduction across the molecule, for a tip positioned above the molecular center, as a function of the number *N* of molecular sites. For off resonance conditions, Fig 6a, the dependence on *N* saturates quickly, within 2-4 segment lengths.[25] When transmission occurs close to resonance (as can be achieved in the model by setting $E_M$ to a value close to zero or by increasing $V_M$ to a value close to 0.25eV that will bring molecular levels close to resonance with the leads Fermi energy) the full extended molecular resonance orbital contribute to the transmission, and the long range dependence on *N* seen in Fig. 6b reflects the sensitivity of this orbital to the value of *N*. Furthermore, the oscillatory dependence on *N* seen in Fig. 6b reflects the structure of the resonant molecular level supported by a given molecular length, in particular the amplitude of this level at the site below the tip (see also discussion of the structure observed in Fig. 7).

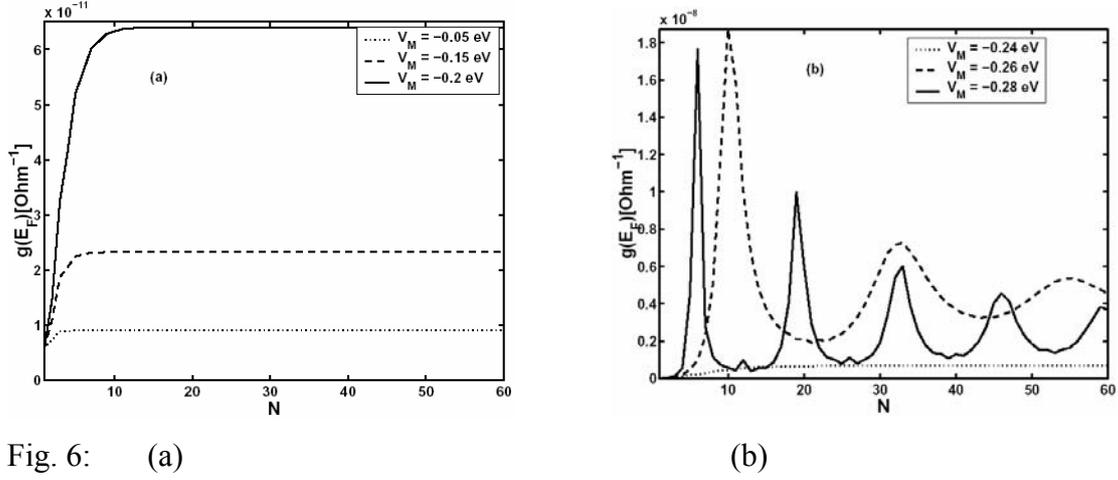

Fig. 6:     (a)                                              (b)

The zero-bias conduction displayed as a function of chain length expressed by the number of sites *N*. (a) Non resonance situations: Dotted line $V_M$=-0.05 eV. Dashed line, $V_M$=-0.15eV. Full line $V_M$=-0.2eV. (b) Dotted line: $V_M$=-0.24 eV, Dashed line: $V_M$=-0.26 eV, Full line: $V_M$=-0.28 eV.

The most important aspect of conduction across a molecule in configuration *b* is its position dependence as observed by scanning the tip along the molecule. The tip–substrate voltage is kept constant, while the current or the tip-height above the substrate are monitored as functions of the tip position along the molecule. Figs 7-9 show the conductance as a function of tip position above our model molecule in configuration *b*. Fig. 7 shows results obtained from the model and parameters of Section 2 with *N*=20 and with the intersite coupling varying to show the effect of site-site connectivity on the image structure. The site structure is seen for values of $V_M$ ($\leq 0.2\text{eV}$) that correspond to non-resonance conditions, but is largely lost (for the current choice of parameters) when $V_M$ is such ($\geq 0.25\text{eV}$) that the transmission involves a delocalized resonant molecular level. Another structure, not related to the site structure, is seen to develop in the resonance regime.

Structure as a function of tip position can results from two reasons. First, naturally, the effective tip-substrate coupling through the molecule changes as the tip moves from a position above a molecular site to a position between sites. This results in a structure dominated by the molecular intersite distance as a characteristic length. Secondly, the molecular wavefunction (obtained by diagonalizing the molecular

Hamiltonian, has an energy dependent spatial structure that can be understood if we think of our molecular model as a discretized box, with the ground states having no nodes on the molecular axis, the first excited state having one node, etc. When such a molecular state comes into resonance with the leads Fermi energy, this structure is reflected in the conduction image. Such structure is seen superimposed on the site induced structure when using $V_M > 0.25 eV$, e.g. the $V_M$ = 0.5eV case of Fig. 7. Such behavior was seen experimentally in STM images of carbon nanotubes.[26] For another theoretical demonstration of this effect see Ref. 27. We note in passing that the structure seen is Fig. 2 more likely belongs to the first kind, i.e. reflects the molecular morphology, since it does not change with the applied voltage in particular when feedback control is applied in order to keep the current constant.

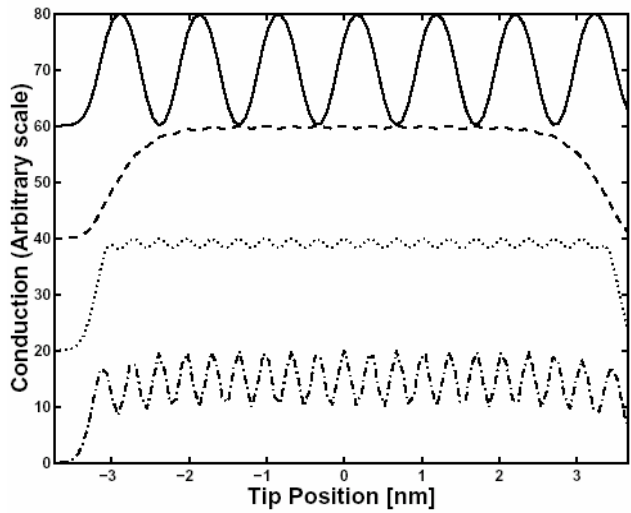

Fig. 7

The zero-bias conduction in configuration *b* displayed against the tip distance from the molecular chain center for a 20 site molecule. Full, dashed, dotted and dashed-dotted lines correspond to $V_M$=-0.5, -0.25, -0.2, and 0.0 eV, respectively. The different lines were scaled to a maximum height of 20 above the g=0 baseline and shifted vertically so as not to overlap.

Adsorption of long molecular chain on surfaces may result in non-ideal structures either because of defects on the original substrate surface or because of local reconstruction affected by the adsorption process. Fig 8 depicts results obtained on one theoretical manifestation of this effect. The system studied is similar to that of Fig. 7, except that a small disorder has been introduced to the site energies. This was done by taking $E_n = E_M + \delta E_n$, where $\delta E_n$ was sampled from a Gaussian distribution characterized by $\langle \delta E_n \rangle = 0$ and $\langle |\delta E_n|^2 \rangle^{1/2} = 0.01 eV$ (as before $E_M$=0.5eV).[29] The irregular structure superimposed on the atomic structure is reminiscent of what is seen in Fig. 2b, even though an uncorrelated Gaussian disorder is probably a poor model of the actual defect distribution in such systems.

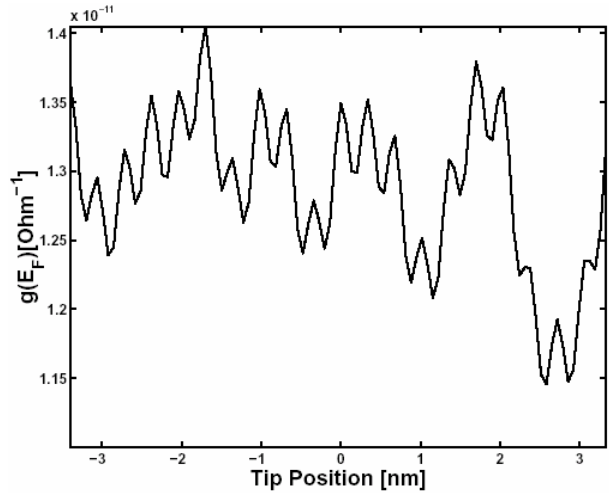

Fig. 8. Same as Fig. 7, using standard parameters (see Sect 2) with $N$=100, except that a small Gaussian noise is added to the zero order site energies (see text for detailes).

Finally, Fig 9 shows the same theoretical STM images obtained in the presence of thermal interactions. Here, for simplicity, we do not calculate the conduction but the transmission coefficient obtained from the formalism described in Sect. 3 and in Ref. 20. Also, to render the calculation numerically efficient we have used relatively short chains with $N$=7, so Fig. 9 shows the corresponding end effects which do not, however change the behavior near the chain center. The strength of the thermal interaction is measured by the parameter $C_T$ of Eq. (21). In this regime, conduction along molecular chains longer than $N$~3-5 is dominated by intersite hopping., however the conduction across the molecule (configuration *b*) is essentially an $N$=1 transmission where coherent effects may still dominate except perhaps very close to resonance. From Fig 9 we see that even in this case the transmission is increased with the strength of the thermal interaction. The most important observation however is that at room temperature, interaction of the molecule with the thermal environment does not affect the overall quality, resolution or shape of the transmission scan.[30]

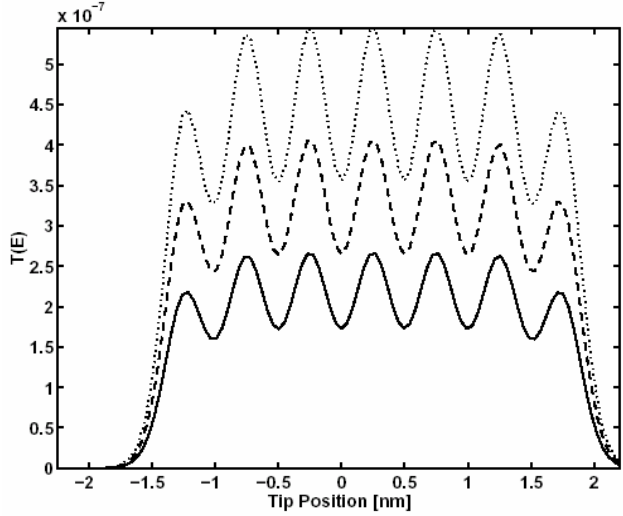

Fig. 9. The transmission probability, $T(E_F) = \int dE_f \mathcal{T}(E_F, R_f)$, in configuration $b$ for the thermal interaction model of Section 3. Dotted line, $C_T = 0$. Dashed line, $C_T = 0.05$. Full line, $C_T = 0.1$.

## 5. Conclusions

Tight binding models have been very useful in investigating basic generic features associated with electrical conduction through molecular bridges. In spite of their oversimplified nature they can account for several fundamental aspects of molecular conduction junction such as conduction gaps, molecular length dependence, resonance and non-resonance behavior, disorder effects and thermal interactions. In this paper we have used such model to characterize different modes of behavior in conduction across molecular chains as observed in STM images of flatly adsorbed molecular chains. While the theory describing conduction is similar for both types of experiments, the different configuration used put emphasis on different aspects of their behavior. In particular, the current-voltage characteristic is the principal observable in the first type of experiments, the position dependence of conduction (or a related quantity) is the principal observable in the other. In principle, carrying out both type of experiments on the same molecular wire can provide a useful consistency check on the theory used to interpret such experiments, since the molecular Hamiltonian (and the corresponding molecular electronic structure) affecting both are the same.

Focusing on the STM signal obtained from our model we have shown that (a) choosing reasonable molecular parameters that lead to conduction along the molecule in the physically reasonable range also results in reasonable calculated STM currents. This is an important validity check on our oversimplified model. (b) The spatial structure of STM images reflects the site structure of the molecular bridge, but in addition may show the spatial structure of molecular wavefunctions that satisfy resonance tunneling conditions. (c) The scanning image is very sensitive to local disorder that may change the local energies and tunneling barriers. (d) Within the model studied, thermal interactions were found to modestly affect the overall transmission probability, therefore the observed tunneling current, however such interactions, in a reasonable parameter range seem not to affect the overall appearance of the image or its resolution. We note that to the best of our knowledge this is the first time that thermal relaxation effects on STM images have been addressed. (e) As an experimental technique to monitor molecular electronic transport properties, the STM configuration, although more sensitive to the junction parameters, enables to overcome the problem of structural defects and impurities in the molecules that may block the current in the leads configuration.

As said repeatedly above, the tight binding model used in this work is grossly oversimplified. Still, the observations made above are general enough in nature to remain valid in more realistic models. Further progress in exploring the relationship between conduction along and across molecular wires can be achieved by addressing both processes by a suitable ab-initio calculation. The fact that the same molecular structure enters in both processes does have the potential to provide an important consistency check on this calculations, or to use results obtained from one type of experiment to infer about the other. It will be interesting to explore such possibilities in the future.

**Acknowledgements:** Our research is supported by the Israel Science Foundation, the U.S.-Israel Binational Science Foundation and the Volkswagen Foundation (AN); by the ISF, FIRST, GIF and IST-FET (DP) and by the EU under grant IST-2001-38951 (GC). DP and HC are grateful to Li Zhu for assistance and useful discussions. This paper is dedicated to Professor Joshua Jortner, who has set the stage for many of the ideas and concepts that have shaped this field, on his 70$^{th}$ birthday.